\documentclass[aps,pra,twocolumn,superscriptaddress]{revtex4-2}
\usepackage[utf8]{inputenc}
\usepackage[english]{babel}
\usepackage{xcolor}
\usepackage{graphicx}
\usepackage[colorlinks,citecolor=blue,urlcolor=blue,linkcolor=blue]{hyperref}
\usepackage{mathtools}
\usepackage{amsmath}
\usepackage{amssymb}
\usepackage{braket}
\usepackage{tikz}

\newcommand{\freiburg}{Physikalisches Institut, Albert-Ludwigs-Universität Freiburg,\\ Hermann-Herder-Straße 3, D-79104 Freiburg, Germany}
\newcommand{\eucor}{EUCOR Centre for Quantum Science and Quantum Computing, Albert-Ludwigs-Universität Freiburg,\\ Hermann-Herder-Straße 3, D-79104 Freiburg,  Germany}

\begin{document}
\title{Hong-Ou-Mandel interference on a lattice: symmetries and interactions}

\author{Mama Kabir Njoya Mforifoum}
\affiliation{\freiburg}
\affiliation{\eucor}

\author{Andreas Buchleitner}
\affiliation{\freiburg}
\affiliation{\eucor}

\author{Gabriel Dufour}
\affiliation{\freiburg}
\affiliation{\eucor}

\begin{abstract}
We describe the Hong-Ou-Mandel interference of two identical particles evolving on a one-dimensional tight-binding lattice where a potential barrier plays the role of a beam splitter. Careful consideration of the symmetries underlying the two-particle interference effect allows us to reformulate the problem in terms of ordinary wave interference in a Michelson interferometer. This approach is easily generalized to the case where the particles interact, and we compare the resulting analytical predictions for the bunching probability to numerical simulations of the two-particle dynamics. 
\end{abstract}
\maketitle

\section{Introduction}

Toss two coins; the most likely outcome is that one will show heads and the other tails.
However, if you perform this experiment with two indistinguishable photons simultaneously impinging on opposite sides of a $50$-$50$ beam splitter, they will always exit on the same side, i.e. either both will show ``heads'' or both will show ``tails''.  This intriguing two-particle-interference effect was first demonstrated in the  1987 experiment of Hong, Ou and Mandel (HOM) \cite{hong1987measurement}.
While our understanding of \emph{single-particle} interference is supported by our experience with classical waves, it is often claimed that such \emph{many-particle} interference has no classical counterpart.
Indeed, the cancellation of amplitudes associated with coincidence events (one particle in each output) in the HOM experiment, although it can be derived in a few lines (see e.g. \cite{bouchard2021two-photon}), remains at first glance an abstract mathematical result.

Notwithstanding, the HOM effect has become a standard tool in quantum optics laboratories, with a wide range of applications, including photon-source characterization, 
precision measurements  
and quantum communication, 
 to cite only a few (we refer the reader to \cite{bouchard2021two-photon} for a recent review).
 Moreover, since beam splitters can be arranged into larger interferometers \cite{reck1994experimental,mayer2011counting}, HOM interference is the basic building block for more complex interference scenarios, e.g.
 multiport interferometers displaying completely destructive interference
\cite{lim2005generalized,tichy2010zero-transmission,tichy2012many-particle,crespi2015suppression,crespi2016suppression,dittel2017many-body,viggianiello2018experimental,dittel2018totally,dittel2018totally-1}
and optical quantum computing schemes \cite{knill2001scheme,kok2007linear,aaronson2013computational}.

In our present contribution, we let two particles evolve on a one-dimensional (1D) lattice, such that we can watch their interference unfold in space and time. Furthermore, by mapping the two-particle 1D onto a single-particle 2D problem, we provide a simple and visual interpretation of HOM interference as ordinary wave interference.
In addition to offering a new perspective on a fundamental phenomenon,
our approach allows to naturally include interactions between the particles.
Indeed, many-particle interference is by no means an exclusive feature of non-interacting photons and  the HOM effect has also been demonstrated with other types of --- potentially interacting --- particles, e.g. with cold bosonic atoms \cite{lahini_quantum_2012,kaufman2014two-particle,preiss_strongly_2015,lopes2015atomic,robens2016quantum,kaufman2018chapter}
and, in its fermionic variant, with electrons in mesoscopic devices \cite{bocquillon2013coherence}. 
Actually, interactions and many-particle interference are arguably the main ingredients behind the complexity of many-body physics, from condensed matter systems to cold atomic gases.  
Understanding the effect of interactions in the HOM experiment \cite{andersson1999quantum,longo2012hong-ou-mandel,gertjerenken2015effects,mullin2015interference,dufour2017many-particle,yannouleas2019interference} thus paves the way to a more general comprehension of the interplay of  interference and interactions in these systems \cite{brunner2018signatures,dufour2020many-body,brunner_manybody_2023}.

Our interference scenario is the following: two particles are initially placed on opposite sides of a potential barrier in a 1D tight-binding lattice, as might for example be realized  with cold atoms in a standing laser wave \cite{bloch2008many-body,preiss2015strongly}.
The particles are launched towards each other and meet at the potential barrier, which plays the role of a beam splitter (see also \cite{longo2012hong-ou-mandel,banchi2015perfect,compagno2015toolbox} for similar approaches). We then examine the probability that both particles exit on the same side of the barrier, i.e. that they bunch, depending on the barrier height, the initial quasimomentum of the particles, their interaction strength and quantum statistics.

 We start by briefly recalling the scattering coefficients of a single particle on a potential barrier and show under which conditions it can act as a $50$-$50$ beam splitter.
 Then, for two particles, we make use of the Hamiltonian's invariance under particle exchange and of the lattice's mirror symmetry to reformulate the problem in terms of the dynamics of a single quantum object in a 2D wedge billiard whose boundary conditions are fixed by the symmetries. This geometry is equivalent to that of a Michelson interferometer, providing us with  an intuitive way of evaluating the bunching and coincidence probabilities in terms of single-particle transmission and reflection coefficients.
 In the non-interacting case, we recover the expected HOM behaviour
 and show how one can switch between bosonic and fermionic interference by tailoring the initial state. Contact interactions are then introduced by a simple modification of the boundary conditions, elucidating their effect on the interference signal. 
  We compare our analytical predictions for the bunching probability to numerical simulations of wavepackets evolving on a finite lattice
  in both non-interacting and interacting scenarios. In the latter case, we discuss the formation of bound states of the two particles, or of one particle on the barrier, which slightly limit the validity of our analytical expressions.

\section{Model}

\subsection{Single particle}

\begin{figure}
	\centering
		\includegraphics[width=0.45\textwidth]{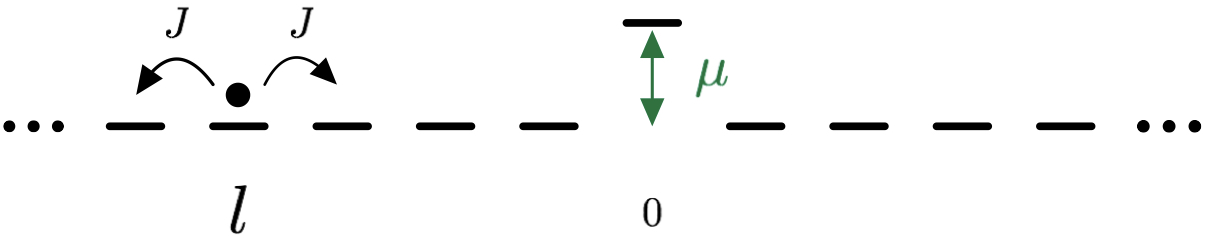}
	\caption{Sketch of the 1D lattice with nearest-neighbor tunnel coupling $J$ and a central potential barrier of height $\mu$. }
	\label{fig:lattice}
\end{figure}

We consider a $1D$ tight-binding lattice in which we incorporate a potential barrier which will play the role of a beam splitter \cite{longo2012hong-ou-mandel,compagno2015toolbox}, as illustrated in Fig \ref{fig:lattice}. The single-particle Hamiltonian reads
\begin{equation}\label{H1p}
	H_1 = -J  \sum_{l}\Big( | l \rangle \langle l+1 | + | l+1 \rangle \langle l | \Big)  + \mu |0 \rangle \langle 0 |~,
\end{equation}
where $|l \rangle$ is the state of the particle localized on the $l^\text{th}$ lattice site, $J$ is the tunnel coupling between neighbouring sites, and $\mu$ the height of the potential barrier located on site $l=0$. Unless otherwise stated, the lattice is assumed to be infinite, such that the site index $l$ runs over the integers, $l\in\mathbb{Z}$. Expanding the quantum state in the site basis, $\ket{\psi}=\sum_l \psi(l)\ket{l}$, the Schrödinger equation (SE) $H_1\ket{\psi}=E\ket{\psi}$ reads
\begin{equation}\label{SE1p}
	-J\left[ \psi(l-1)+\psi(l+1)\right] +\mu \delta_{l,0}\psi(l)  = E \psi(l)~.
\end{equation}
Away from the potential barrier ($l\neq 0$), the SE accepts counterpropagating plane waves with quasimomentum $\pm k$, 
$\psi(l)=\exp(\pm i k l)$, as degenerate solutions of energy $E=-2J\cos k$.

The SE \eqref{SE1p} evaluated at $l=0$ allows to connect the incident, reflected and transmitted waves with quasimomentum $\pm k$, yielding the  transmission and reflection coefficients \cite{longo2012hong-ou-mandel,njoyaPhD}
\begin{align}\label{t}
T &= \frac{2 i J \sin k}{2 i J  \sin k  -\mu}\\
 \text{and} \quad
\label{r}
R &= \frac{\mu }{2 i J  \sin k - \mu}~.
\end{align}
In particular, for the potential barrier to act as a $50$-$50$ beam splitter, i.e.  with transmission and refection probabilities $|T|^2 = |R |^2 = \frac{1}{2}$, we find  the following condition
\begin{equation}
\mu_{50\text{-}50} = \pm 2 J \sin k~.
\end{equation}

\subsection{Two particles -- Symmetries}

\begin{figure}
	\centering
	\includegraphics[width=0.5\textwidth,trim=4cm 7cm 3cm 1cm,clip]{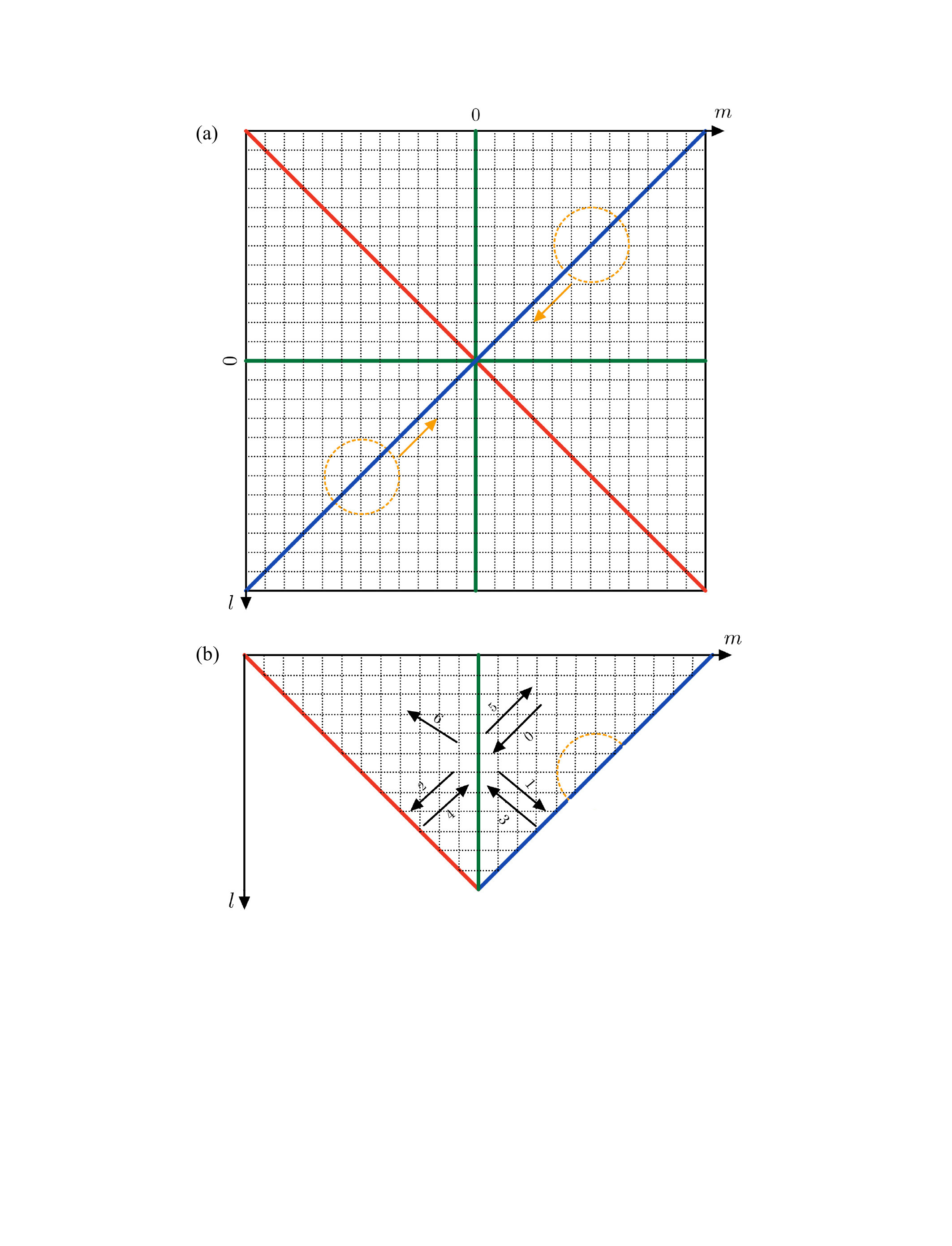}
	\caption{(a) Configuration space of two particles on a 1D lattice. The horizontal  (vertical) axis corresponds to the position $l$ ($m$) of the first (second) particle. The green lines indicate configurations where one of the particles is on the potential barrier ($l=0$ or $m=0$). The red diagonal marks configurations where both particles are located on the same site ($l=m$), the blue antidiagonal those where the particles are placed symmetrically with respect to the barrier ($l=-m$). The yellow circle and arrow indicate the position and quasimomentum of an initial state of identical particles.
		(b) Reduced configuration space for states with well-defined parity under the transformations $E$ and $D$, obtained by folding the full configuration space along its diagonal and antidiagonal. The ensuing boundaries are referred to as the $E$ (red line) and $D$ mirrors (blue line), respectively. The numbered arrows indicate the various propagation directions that are coupled by scattering on the potential barrier and on the boundaries.}
	\label{fig:confspace}
\end{figure}

We write the state of two particles as
$\ket{\Psi}=\sum_{l,m}\Psi(l,m)\ket{l,m}$
in the tensor-product basis $\ket{l,m}=\ket{l}\otimes\ket{m}$, where $l$ is the position of the first, and $m$ that of the second particle.
The wavefunction $\Psi(l,m)$ is thus defined on the 2D configuration space depicted in Fig.~\ref{fig:confspace}~(a), such that the two-particle evolution can equivalently be seen as that of a single quantum object on a 2D square lattice.

In the presence of contact interactions, the two-particle Hamiltonian reads
\begin{equation}\label{H2p}
H_2 = H_1 \otimes \mathbb{I} + \mathbb{I} \otimes  H_1  + U \sum_{l} | l, l \rangle \langle l, l |~,
\end{equation}
with $H_1$ the single particle Hamiltonian of Eq.~\eqref{H1p}, and $U$ the energy cost for two particles to occupy the same site.
With the goal to simplify the analysis of the Hamiltonian $H_2$, we consider the following symmetry operations: particle exchange $E$  and lattice parity $P$, which act on basis states as
\begin{align}\label{ep}
E | l,m \rangle &=   | m,l \rangle\\
 \text{and} \quad P | l,m \rangle &=  | -l,-m \rangle~,
\end{align}
respectively.
In the two-particle configuration space sketched in  Fig.~\ref{fig:confspace}~(a),
parity $P$ corresponds to inversion through the central point and exchange $E$ to reflection across the diagonal (red line).
It is convenient to also define the joint operation $D=EP=PE$ (note that $E$ and $P$ commute), which is given by reflection across the antidiagonal (blue line).
We can thus decompose the two-particle state $\ket{\Psi}$ into components $\ket{\Psi_{\epsilon,\delta}}$ with definite symmetry properties under reflection on the two diagonals of the configuration space:
\begin{align}
E |\Psi_{\epsilon,\delta} \rangle &=  \epsilon | \Psi_{\epsilon,\delta} \rangle\\
\text{and} \quad D |\Psi_{\epsilon,\delta} \rangle &= \delta | \Psi_{\epsilon,\delta} \rangle~, 
\end{align}
where $\epsilon$ and $\delta$ can take the values $\pm1$. For two identical bosons (respectively fermions), only components with $\epsilon=1$ (respectively $\epsilon=-1$) are allowed.

Given that $[H_2,E]=[H_2,D]=[E,D]=0$, the various components $\ket{\Psi_{\epsilon,\delta}}$ are not coupled by the dynamics and can be treated independently. For given values of $\epsilon$ and $\delta$, we can therefore fold the two-particle configuration space along the symmetry axes of $E$ (red line) and $D$ (blue line), as shown in  Fig.~\ref{fig:confspace}~(b), since the state is fully specified by fixing the wavefunction on the resulting reduced configuration space.
The two diagonals of the original configuration space therefore become boundaries of the reduced configuration space, where $\epsilon$- and $\delta$-dependent boundary conditions apply, as we will detail further down. 
In the following, we refer to these boundaries as $E$ mirror (red line) and $D$ mirror (blue line), respectively. For now, we simply denote by $e^{i\phi_\epsilon}$ and $e^{i\varphi_\delta}$ the phases acquired by the component $|\Psi_{\epsilon,\delta} \rangle$ upon normal reflection on these mirrors.

With these considerations, the two-particle dynamics is mapped to that of a single object bouncing between two mirrors intersecting at a right angle, in the presence of a potential barrier bisecting that angle [green line in  Fig.~\ref{fig:confspace}~(b)]. This correspondence will allow us to treat  two-particle interference as ordinary (single-particle) interference in this ``corner reflector''.

\section{Results}

\subsection{Coincidence and bunching probabilities}

We consider two indistinguishable particles (identical bosons, $\epsilon=1$, or fermions, $\epsilon=-1$) initially localized on opposite sides of the barrier and with opposite quasimomenta. This yields an even state under reflection $D$, i.e., a state with $\delta=1$. The positions and quasimomenta of the initial two-body wavefunction are represented by the yellow circles and arrows in Fig.~\ref{fig:confspace}~(a).
The state then evolves with the Hamiltonian $H_2$, such that the particles scatter against the potential barrier and each other. Once the wavefunction has again moved away from the central region, we can evaluate the bunching and coincidence probabilities.
The bunching probability $P^\mathrm{b}$ is the probability of finding both particles on the same side of the beam splitter, i.e. in the upper left or lower right quadrants of the configuration space sketched in Fig.~\ref{fig:confspace}~(a). Its complement to one, the coincidence probability $P^\mathrm{c}=1-P^\mathrm{b}$, is the probability of finding exactly one particle on each side of the beam splitter [lower left and upper right quadrants in Fig.~\ref{fig:confspace}~(a)].

\begin{figure}
	\centering
	\includegraphics[width=0.8\columnwidth]{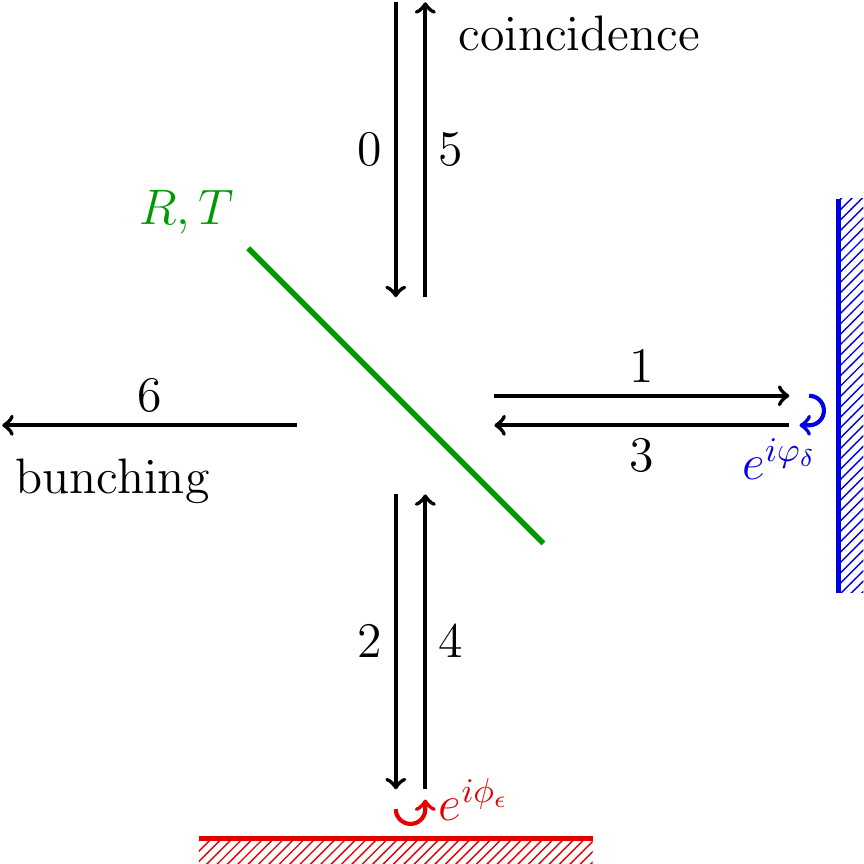}
%
%
	\caption{\label{fig:interference} Equivalent Michelson interferometer 
		obtained by viewing Hong-Ou-Mandel interference in the reduced 2D configuration space of Fig.~\ref{fig:confspace}~(b). The numbered arrows correspond to the propagation directions also indicated there. The scattering coefficients associated with scattering on the beam splitter (green) and on the $E$ and $D$ mirrors (red and blue, respectively) are indicated.}
\end{figure}

In the reduced configuration space shown in Fig.~\ref{fig:confspace}~(b), the initial position is represented by the yellow half-circle and the initial quasimomentum by the arrow numbered 0. 
Possible propagation directions after scattering on the barrier and boundaries are represented by the other numbered arrows. Coincidence and bunching events are associated with the triangular regions respectively to the right and left of the green line depicting the barrier, i.e. to the final propagation directions indicated by arrows 5 and 6, respectively.
In the absence of a barrier ($\mu=0$, $T=1$, $R=0$), the incident wavepacket is reflected back from the $E$ mirror, such that only coincidence events are observed. For finite values of the barrier height $\mu$, however, the wavepacket is split into two: one component  continues
towards the $E$ mirror while the other is diverted towards the $D$ mirror. After normal reflection on the respective mirrors, both components recombine on the barrier. The two-particle Hong-Ou-Mandel interference is thus mapped to single-particle interference in a Michelson interferometer, as sketched in Fig.~\ref{fig:interference}, with the Michelson's two output ports corresponding to bunching and coincidence events, respectively.

 The interfering pathways in the interferometer can be described by a sequence of propagation directions, as given by the numbered arrows in  Figs.~\ref{fig:confspace}~(b) and~\ref{fig:interference}. 
Paths $0 \rightarrow 2 \rightarrow 4 \rightarrow 6$ and $0\rightarrow 1 \rightarrow 3\rightarrow 6$ lead to bunching events. The former is associated with the amplitude  $T e^{i\phi_\epsilon} R$ and the latter with the amplitude $R e^{i\varphi_\delta} T$, such that the bunching probability reads
 \begin{equation}\label{Pb}
 	P^\mathrm{b}_{\epsilon,\delta}=|e^{i\phi_\epsilon}+e^{i\varphi_\delta}|^2 |R T|^2~,
 \end{equation}
where we have made the dependence of the bunching probability on $\epsilon$ and $\delta$ explicit.
Coincidence events result from the interference of the paths $0 \rightarrow 1 \rightarrow 3 \rightarrow 5$  and $0 \rightarrow 2 \rightarrow 4 \rightarrow 5$,  with contributions $R e^{i\varphi_\delta}R $ and $T e^{i\phi_\epsilon} T$, respectively. As a consequence, the coincidence probability reads 
 \begin{equation}
P^\mathrm{c}_{\epsilon,\delta}=|e^{i\phi_\epsilon}T^2+e^{i\varphi_\delta}R^2|^2~. 
\end{equation}
These two probabilities are seen to sum to one, given that scattering at the barrier
is unitary ($|R|^2+|T|^2=1$ and $RT^*+R^*T=0$). From now on, we therefore focus on the bunching probability $P^\mathrm{b}_{\epsilon,\delta}$.

\subsection{Non-interacting particles}

To determine the reflection phase  $\phi_\epsilon$ on the $E$ mirror, it is useful to briefly come back to the original 2D configuration space [Fig.~\ref{fig:confspace}~(a)] and consider a wavepacket impinging normally on the main diagonal (red line). In the absence of interactions ($U=0$), nothing distinguishes the diagonal configurations $\ket{l,l}$ from the others ($\ket{l,m}$, with  $l\neq m$), such that the diagonal is transparent for the incoming wavepacket. 
If we now consider a state which is (anti)symmetric with respect to the diagonal, it is simply replaced by its (sign-flipped) mirror image as it reaches the diagonal, i.e. we have $e^{i\phi_\epsilon}=\epsilon$. Following a similar reasoning, but this time irrespective of the interaction strength,  we also have $e^{i\varphi_\delta}=\delta$.
With Eqs.~\eqref{t}, \eqref{r} and \eqref{Pb}, it follows that, in the non-interacting case, 
\begin{align}\label{bunchnonint}
P^\mathrm{b}_{\epsilon,\delta} &= |  \epsilon + \delta|^2 |T R|^2\\
&=|  \epsilon + \delta|^2 \frac{4J^2 \mu^2 \sin^2 k}{(4J^2 \sin^2 k+\mu^2)^2}~.\label{bunchnonint2}
\end{align}
In particular, this expression reproduces well-known results \cite{loudon1998fermion,longo2012hong-ou-mandel,compagno2015toolbox} for the specific choice $\delta=1$: while the bunching probability vanishes independently of the beam splitter parameters for fermions ($P^\mathrm{b}_{-+} =0$), the bunching probability of bosons reads $P^\mathrm{b}_{++} =4|T R|^2$ and reaches 100\%  for a balanced beam-splitter  ($|T|^2=|R|^2 =\frac{1}{2}$).


We now compare these predictions with numerical simulations on a finite lattice with $L$ sites (we take $L$ odd and $ \frac{1-L}{2} \leq l \leq \frac{L-1}{2}$). The particles are prepared in Gaussian wavepackets with central quasimomenta $k_l=-k_m=k> 0$, central positions $l=-m=c<0$ and width $\sigma$:
\begin{align}\label{init}
	&\ket{\Psi_0}= \sum_{l,m} \text{G}_{k,c,\sigma}(l) \text{G}_{-k,-c,\sigma}(m)\ket{l,m}~,\\
	\text{with} \quad &
	\text{G}_{k,c,\sigma}(l)=  \exp\left(- \frac{(l-c)^2}{4\sigma^2} +i k l \right)~.
\end{align}
The state is then symmetrized according to the particles' quantum statistics, i.e. we apply $\mathbb{I}+\epsilon E$, before normalizing.

The initial distance $|c|$ of the particles from the barrier must be large enough compared to the width $\sigma$ of the wavepackets, such that the two particles initially have no overlap with each other, nor with the central barrier. 
On the other hand, $\sigma$ should be large enough compared to $1$, such that the quasimomentum distribution of the Gaussians is peaked around $\pm k$ and the wavepackets propagate at the group velocity   $v_g=\hbar^{-1}\partial E/\partial k=2 \hbar^{-1} J \sin k$ without excessive dispersion.
Under these conditions, the wavepackets reach the centre of the lattice after an evolution time $|c|/v_g$ and reach the borders of the system  at $t = (|c|+L/2)/v_g$, at which time we evaluate the bunching probability as
\begin{equation}\label{Pbnum}
	P^\mathrm{b}_{\epsilon,\delta}=\sum_{l,m\, :\, lm>0}|\Psi(l,m;t)|^2~,
\end{equation}
with $\Psi(l,m;t)$ the state's amplitude on basis state $\ket{l,m}$ after an evolution time $t$.
This approach is applicable away from the quasimomenta $0$ and  $\pm \pi$, since in these limits the group velocity vanishes and the dispersion is maximal, preventing simulation on a finite lattice.

\begin{figure}
	\centering
		\includegraphics[width=0.37\textwidth]{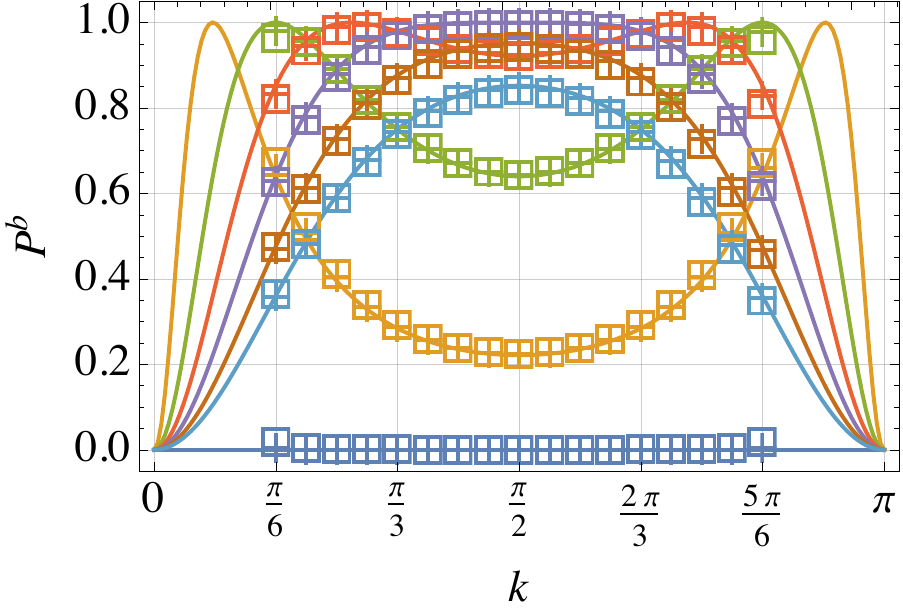}
		\parbox[b][3.85cm][t]{0.1\textwidth}{\ \ \includegraphics[width=0.08\textwidth]{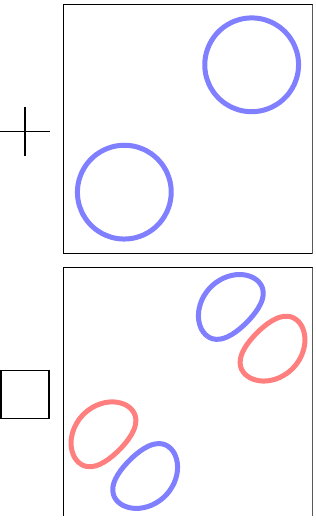}}
		
		\includegraphics[width=0.45\textwidth]{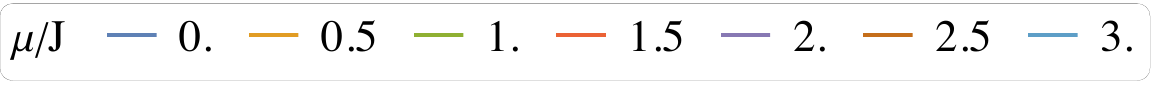}
	\caption{Bunching probability for non-interacting particles ($U=0$), as a function of the relative quasimomentum $k$, for various values of the barrier height in units of the tunnelling strength $\mu/J$ (colour code in the legend). The full lines show the analytical predictions $P^\mathrm{b}_{++}=P^\mathrm{b}_{--}$ for bosons in a $D$-even state, or for fermions in a $D$-odd state. The cross and square markers, respectively, show the results of the numerical wavepacket propagation for bosons in initial state \eqref{init}, and for fermions in initial state \eqref{initodd}. These initial states are sketched in the right panel. In both cases, $L=61$, $|c|=15$, and $\sigma=5$.}
	\label{fig:nonint}
\end{figure}

In our simulations, we take a lattice of size $L = 61$, initial states of width $\sigma=5$ starting at a distance $|c|=15$ from the lattice centre, and vary the values of the quasimomentum $\pi/6 \leq k \leq 5\pi/6$ and of the barrier height $0 \leq \mu \leq 3 J$.
As shown in Fig.~\ref{fig:nonint}, for this range of parameters,  we observe a good agreement between the numerical simulation with finite wavepackets and the analytical prediction from Eq.~\eqref{bunchnonint},  which uses the reflection and transmission coefficients for plane waves [Eqs.~\eqref{r} and \eqref{t}].

In the expression \eqref{bunchnonint} for the bunching probability, it appears that the behaviour of bosonic and fermionic particles can be swapped by changing the parity of the initial state from $\delta=+1$ to $\delta=-1$.
This can for example be realized by choosing the initial state (before symmetrization and normalization)
\begin{align}\label{initodd}
	\ket{\Psi_0}&= \sum_{l,m} (l+m)\text{G}_{k,c,\sigma}(l) \text{G}_{-k,-c,\sigma}(m)\ket{l,m}~.
\end{align}
Note that, in contrast to the product state \eqref{init}, due to the weights $(l+m)$ in the expansion \eqref{initodd}, the positions of the two particles are now entangled.
The bunching probability of fermions ($\epsilon=-1$) prepared in the $\delta=-1$ state \eqref{initodd} is shown in Fig.~\ref{fig:nonint} and is seen to match that of bosons ($\epsilon=1$) prepared in the $\delta=1$ state \eqref{init}, up to small finite-size effects. This is clear from our derivation, since the interference depends only on the difference between the phases acquired upon reflection on the $E$ and $D$ mirrors, which is the same for both states \eqref{init} and \eqref{initodd}, irrespective of their distinct decompositions in the $\{\ket{l,m}\}$ basis.

For states with indefinite symmetry $\ket{\Psi}=\sum_{\epsilon,\delta} c_{\epsilon,\delta}\ket{\Psi_{\epsilon,\delta}}$, each component $\ket{\Psi_{\epsilon,\delta}}$ evolves independently, and  the total bunching probability is obtained as the weighted sum of the bunching probabilities of the individual (mutually orthogonal) components:  
$P^\mathrm{b}=\sum_{\epsilon,\delta} |c_{\epsilon,\delta}|^2  P^\mathrm{b}_{\epsilon,\delta}$.
In particular, we see that the classical bunching probability $P^\mathrm{b}= 2|RT|^2$ can be achieved in two ways:
either by sufficiently delaying one particle with respect to the other, modifying its central position and/or quasimomentum until the weights of the $\delta=1$ and $\delta=-1$ components are balanced, or by leaving out the symmetrization of the initial state, i.e. by considering distinguishable particles, which leads to equal weights on the $\epsilon=1$ and $\epsilon=-1$ sectors.

\subsection{Interacting particles}

In this section, we go back to states with $\delta=1$ and focus on the impact of non-zero contact interaction between the particles ($U\neq 0$). Compared to the non-interacting case, the wavepacket picks up an additional phase upon reflection on the $E$ mirror, because of the interaction between the particles which occurs there.
To determine this phase, we go back to our reasoning at the beginning of the previous section: reflection on the $E$ mirror in the reduced configuration space [Fig. \ref{fig:confspace}~(b)] results 
 from the interference of two processes in the original  configuration space [Fig.~\ref{fig:confspace}~(a)]: the reflection of the  incoming wavepacket on the interaction potential and the transmission of its mirror image on that same potential.
 To determine the corresponding reflection and transmission coefficients,
we consider a wavefunction $\Psi(l,m)=\psi_\mathrm{r}(l - m)$ which depends only on the relative coordinate of the particles, corresponding to motion in the direction orthogonal to the interaction line $l=m$ [red line in Fig.~\ref{fig:confspace}~(a)]. The two-particle SE for $\Psi(l,m)$ is seen to reduce to a single-particle SE for the relative wavefunction $\psi_\mathrm{r}$, with a Hamiltonian of the same form as $H_1$ [Eq.~\eqref{H1p}] but with the replacements $J\to2 J$ and $\mu \to U$.
Therefore, using the results provided in the single-particle case, Eqs.~\eqref{t} and \eqref{r}, the transmission and reflection coefficients on the interaction line
 for a plane wave propagating along the relative coordinate,
 $\Psi(l,m)=\exp(i k (l-m))$, read
 \begin{align}
 	T' &= \frac{4 i J  \sin k}{4 i J \sin k  - U}\\ 
 	\text{and}\quad R' &= \frac{U}{4 i J  \sin k - U}~.
 \end{align} 
Accordingly, the reflection phase on the $E$ mirror reads
\begin{equation}
	e^{i\phi_\epsilon}=R'+\epsilon T'=
	\begin{cases}
		-1  &\text{if } \epsilon=-1 \text{ (fermions)}~,\\
		\frac{4 i J  \sin k +U}{4 i J \sin k  - U} &\text{if } \epsilon=1 \text{ (bosons)}~.
	\end{cases}
\end{equation}




In the case of fermions, the antisymmetry of the wavefunction prevents the particles from interacting and we see no difference to the non-interacting case: the bunching probability vanishes. For bosons, however, the reflection phase $\phi_{\epsilon=1}$ on the $E$ mirror deviates from zero for finite $U$, leading to a bunching probability [compare Eqs.~\eqref{Pb} and \eqref{bunchnonint2}]
\begin{align}\label{bunchint}
	P^\mathrm{b}_{++}(U) &=\frac{16 J^2  \sin ^2k}{16 J^2 \sin^2 k  + U^2}  P^\mathrm{b}_{++}(U=0) \\
	&= \frac{256 J^4 \mu^2 \sin^4 k}{(16 J^2 \sin^2 k  + U^2)(4J^2 \sin^2 k+\mu^2)^2}~. 
\end{align}
Interactions are here seen to always lead to a reduction of the bunching probability, with the non-interacting value simply being multiplied by a function of $U/(J\sin k)$ that is smaller than 1. In particular, in the limit $U\to\infty$, the bunching probability goes to zero, i.e. hard-core bosons behave in the same way as fermions \cite{compagno2015toolbox,gertjerenken2015effects,mullin2015interference}.

\begin{figure}
     (a)
	
		\includegraphics[width=0.45\textwidth]{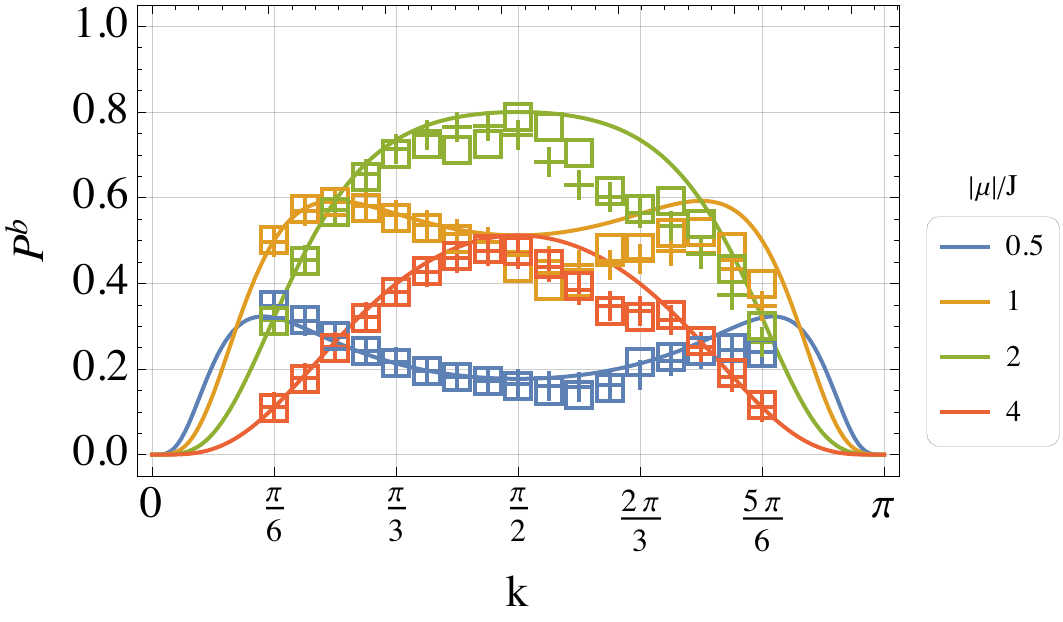}
	
   (b)
	
		\includegraphics[width=0.45\textwidth]{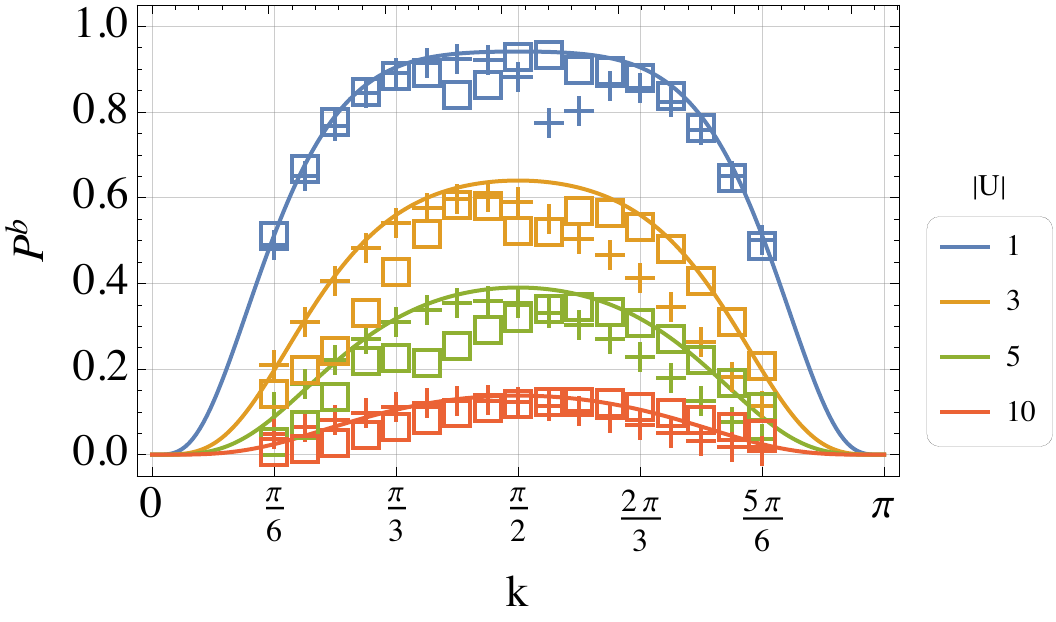}
	\caption{\label{fig:int} 
Bunching probability for interacting particles as a function of the relative quasimomentum $k$, for $U=2J$ and various values of the barrier height $\mu$ [panel (a)], and for $\mu=2J$ and various values of the interaction strength $U$ [panel (b)].
 The full lines show the analytical predictions $P^\mathrm{b}_{++}$ for bosons in a $D$-even state, which are independent of the sign of $\mu$ and $U$. 
  The markers show the results of the numerical wavepacket propagation for bosons in initial state \eqref{init} and positive (crosses) or negative (squares) values of $\mu$ [panel (a)] and $U$ [panel (b)]. In all simulations, $L=61$, $|c|=15$ and $\sigma=5$. We observe localized deviations of the numerical results from the analytical predictions, e.g. for $\mu=U=2J$ and $k\approx 3\pi/4$ [green crosses in panel (a)], which we attribute to the resonant formation of bound states (see main text and state snapshot in Fig.~\ref{fig:bound}).}	
\end{figure}

We now compare these predictions to numerical simulations using the same lattice parameters and initial conditions [Eq.~\eqref{init}] as in the non-interacting case, for various values  of the contact interaction $U$.
As shown in Fig.~\ref{fig:int}~(a), while the overall behaviour of the bunching probability is well captured by the formula Eq.~\eqref{bunchint}, we do observe sizeable deviations of the simulation results from the analytical predictions, which cannot be attributed to finite-size effects alone.
Overall, the analytical prediction tends to overestimate the bunching probability. Moreover, while the formula Eq.~\eqref{bunchint} is invariant under each of the transformations $k\to \pi-k$, $U\to-U$ and  $\mu \to -\mu$, the numerical bunching probabilities are seen to break these symmetries.

These discrepancies can be explained by the formation of single and two-particle bound states, which is neglected in our approach. Indeed, in addition to the plane wave solutions discussed above, Eq.~\eqref{SE1p} also admits solutions with complex quasimomentum $k=\pm i \kappa$ (for $\mu<0$) or $k=\pi \pm i \kappa$ (for $\mu>0$),  with $\kappa>0$,  which decay exponentially away from the barrier:
\begin{align}
	\psi(l)=(-\text{sign}\ \mu)^l \exp(-\kappa |l|)~,
\end{align}
with energy
\begin{align}
	 E_\mathrm{b}=\text{sign}\ \mu \cosh \kappa = \text{sign}\ \mu\ \sqrt{4J^2+\mu^2}~.
\end{align}
Each particle can thus bind to the barrier and---recall the mapping of two interacting particles to a single particle with a barrier discussed at the beginning of this section---the two particles can bind together \cite{valiente2008two-particle,petrosyan2010exotic,boschi2014bound,beggi2018probing}. Bound states of the two particles occur also for repulsive interactions ($U>0$), forming so-called repulsively-bound pairs \cite{winkler2006repulsively}.

\begin{figure}
	\centering
    \includegraphics[width=0.45\textwidth]{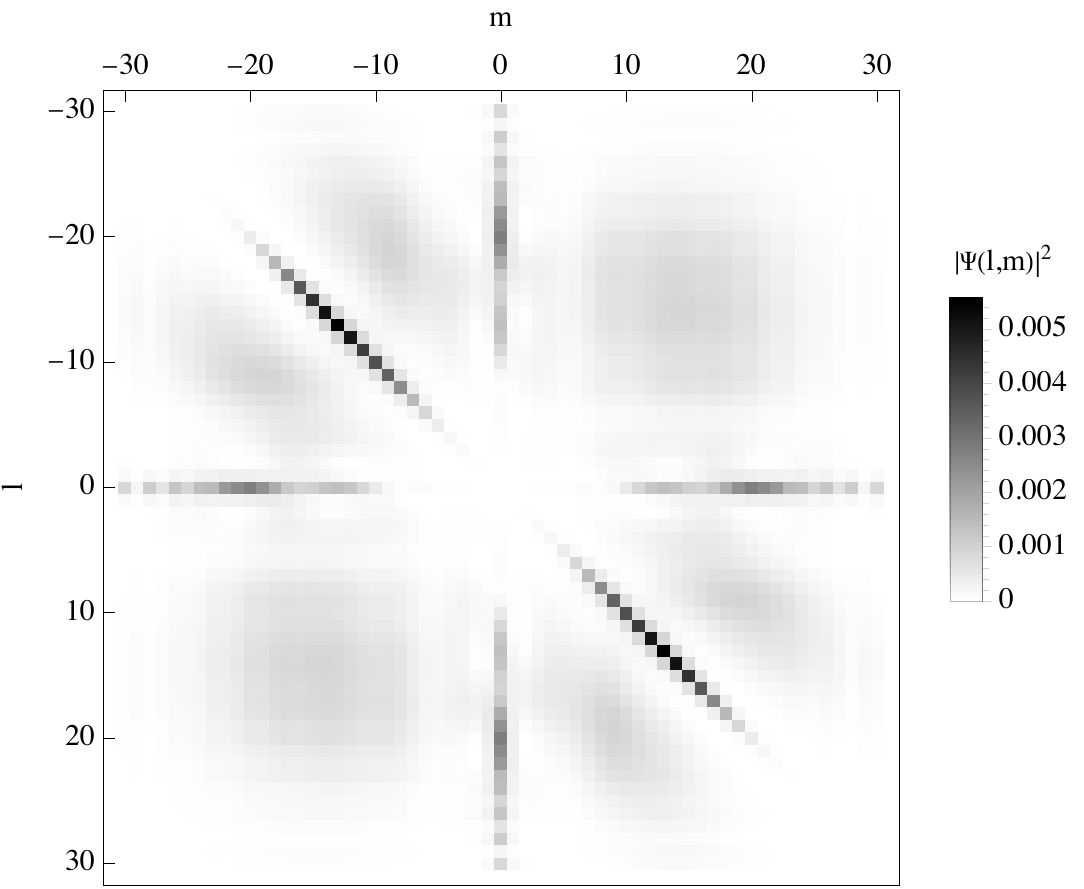}
	\caption{\label{fig:bound}
		Joint probability distribution $|\Psi(l,m;t)|^2$ after an evolution time $t=2c/v_g$ from the initial state \eqref{init}, with $k=3\pi/4$, $|c|=15$, and $\sigma=5$, and Hamiltonian parameters $\mu=U=2J$. Notice the relatively high probability of finding one particle on the barrier or both particles at the same position.	
	}
	
\end{figure}

Such bound states can clearly be seen in the exemplary joint probability distribution $|\Psi(l,m;t)|^2$, shown in Fig.~\ref{fig:bound}, for an initial state \eqref{init} with $k=3\pi/4$, $|c|=15$ and $\sigma=5$ evolving on a lattice of length $L=61$, with parameters $\mu=U=2J$, at time $t=2|c|/v_g$.
The probability densities along the horizontal ($l=0$) and vertical ($m=0$) lines  correspond to states where one of the particles binds to the barrier, while the other one remains in a scattering state \footnote{Note that our numerical evaluation of the bunching probability according to Eq.~\eqref{Pbnum} excludes configurations where one particle is located exactly on the barrier.}. Such states can form since interactions allow energy exchange between the particles. 
Moreover, the potential barrier breaks the translational symmetry of the lattice, and therefore the conservation of the centre-of-mass quasimomentum $k_l+k_m=0$, allowing for the formation of bound states of the two particles, which move away from the barrier together.
In Fig.~\ref{fig:bound}, these states are localized on the diagonal ($l=m$) of the two-particle configuration space and contribute to the bunching probability.

A more detailed analysis of the bunching of interacting particles should take into account the finite cross-section for the formation of such bound states. Numerically, the effect on the bunching probability is seen to be rather moderate on the whole (see Fig.~\ref{fig:int}), although we do observe significant deviations for specific parameter combinations [e.g., $\mu=U=2J$ and $k\approx 3\pi/4$, corresponding to the green crosses in Fig.~\ref{fig:int}~(a) and to the evolved state depicted in Fig.~\ref{fig:bound}], hinting at resonant behaviour.

\section{\label{sec:Conclu}Conclusion}

We analyzed in detail the dynamics of two identical particles impinging on a potential barrier in the centre of a $1D$ lattice. In the non-interacting case, we recovered the well-known Hong-Ou-Mandel interference. The originality of our approach is that the bunching and coincidence probabilities appear as the result of single-particle interference in a 2D wedge which forms a Michelson interferometer. Furthermore, the generalization to the interacting case is almost immediate if one neglects the formation of bound states. Our work thus provides a new perspective on two-particle interference and sets the stage for further studies of many-body interference with interacting particles. For example, in a companion paper \cite{njoya-letter}, we apply this framework to study the Hong-Ou-Mandel interference of composite particles.

\section*{Acknowledgments}
The authors acknowledge support by the state of Baden-W\"urttemberg through bwHPC. MKNM acknowledges support by the German Research Foundation (DFG) through research training groups IRTG 2079 CoCo and RTG 2717 Dyncam.

\bibliographystyle{apsrev4-2} 
\bibliography{LatticeHOM}

\end{document}